\documentclass[final,1p,times,twocolumn,authoryear]{elsarticle}

\usepackage{amssymb}

 \journal{}
\usepackage{amsmath}
\usepackage{graphicx}

\begin{document}

\begin{frontmatter}



\title{Astrocytes mediate analogous memory in a multi-layer neuron-astrocytic network}\author[mymainaddress]{Yuliya Tsybina}
\author[mymainaddress,mysecondaryaddress,myfourthaddress]{Innokentiy Kastalskiy}
\author[mymainaddress]{Mikhail Krivonosov}
\author[mymainaddress,mysevenaddress,myeightaddress]{Alexey Zaikin}
\author[mymainaddress,mythirdaddress,myfourthaddress,myfifthaddress]{Victor Kazantsev}
\author[mymainaddress,mysixaddress]{Alexander Gorban}
\author[mymainaddress,mythirdaddress]{and Susanna Gordleeva\corref{mycorrespondingauthor}}
\cortext[mycorrespondingauthor]{Corresponding author}

\address[mymainaddress]{Scientific and Educational Mathematical Center "Mathematics of Future Technology", Lobachevsky State University of Nizhny Novgorod, Nizhny Novgorod, Russia}
\address[mysecondaryaddress]{Laboratory of Autowave Processes, Institute of Applied Physics of the Russian Academy of Sciences, Nizhny Novgorod, Russia}
\address[mythirdaddress]{Neuroscience and Cognitive Technology Laboratory, Center for Technologies in Robotics and Mechatronics Components, Innopolis University, Innopolis, Russia}
\address[myfourthaddress]{Center for Neurotechnology and Machine Learning, Immanuel Kant Baltic Federal University, Kaliningrad, Russia}
\address[myfifthaddress]{Neuroscience Research Institute, Samara State Medical University, Samara, Russia}
\address[mysixaddress]{Dept of Mathematics, Leicester University, Leicester, UK}
\address[mysevenaddress]{Centre for Analysis of Complex Systems, Sechenov First Moscow State Medical University (Sechenov University), Moscow, Russia}
\address[myeightaddress]{Institute for Women’s Health and Department of Mathematics, University College London, London, UK}

\begin{abstract}

Modeling the neuronal processes underlying short-term working memory remains the focus of many theoretical studies in neuroscience. Here we propose a mathematical model of spiking neuron network (SNN) demonstrating how a piece of information can be maintained as a robust activity pattern for several seconds then completely disappear if no other stimuli come. Such short-term memory traces are preserved due to the activation of astrocytes accompanying the SNN. The astrocytes exhibit calcium transients at a time scale of seconds. These transients further modulate the efficiency of synaptic transmission and, hence, the firing rate of neighboring neurons at diverse timescales through gliotransmitter release. We show how such transients continuously encode frequencies of neuronal discharges and provide robust short-term storage of analogous  information. This kind of short-term memory can keep operative information for seconds, then completely forget it to avoid overlapping with forthcoming patterns. The SNN is inter-connected with the astrocytic layer by local inter-cellular diffusive connections. The astrocytes are activated only when the neighboring neurons fire quite synchronously, e.g. when an information pattern is loaded. For illustration, we took greyscale photos of people's faces where the grey level encoded the level of applied current stimulating the neurons. The astrocyte feedback modulates (facilitates) synaptic transmission by varying the frequency of neuronal firing. We show how arbitrary patterns can be loaded, then stored for a certain interval of time, and retrieved if the appropriate clue pattern is applied to the input.                 

\end{abstract}

\end{frontmatter}

\section{Introduction}

Understanding principles of brain information processing remains one of the main challenges in neuroscience \citep{Chaudhuri2016,Benna2016}. In theory, there is a gap between molecular and cellular levels implementation and its functionality at the cognitive level. Scholars proposed a variety of conceptual, mathematical, and computational models of neuronal networks pretending to implement cognitive functions, such as learning and memory \citep{Hopfield1982, Mongillo2008, Goldman2009, Zenke2015, Lobo2020, Solovyeva2016, Lobov2021,Gorban}. In system neuroscience, memory is a substantially complicated paradigm involving different types and forms. Working memory represents one of these types \citep{Baddeley2012}. Like operative memory in computers, it can store several patterns for operative use for several seconds. After that, some patterns can be selected to be further memorized and the others are completely erased.
Working memory is believed to be ''encoded''  by changes in the strengths of synaptic connections, e.g. synaptic plasticity \citep{Mongillo2008, Hansel2013, Lundqvist2018}. These changes determine, which particular neuronal clusters or signal transmission pathways encoding the information should be memorized. When an appropriate clue is applied, the information can be retrieved in the form of spatio-temporal neuronal firing pattern reproducing original information. In modeling, the design of an adequate mathematical model that can possess both biological plausibility and processing functionality is still an open question \citep{Fiebig2016, Mi2017}. 

Investigations conducted in the last decade reveal more and more aspects related to the implementation of information functions of the CNS. The list of functions performed by the astrocytic cells keeps getting frequently updated and revised \citep{Perea2005, Kimelberg2010, Fields2013, Rusakov2014, LopezHidalgo2014, Vasile2017}. Several studies discuss the role of astrocytes in the perception of sensory stimuli \citep{Lines2020, Stobart2018, Reynolds2019, Chen2012}, spatio-temporal coordination of neural network signaling \citep{Sonoda2018, Gordleeva2019, Kanakov2019}, information processing, and cognitive functions \citep{Oliveira2015, Paukert2014, Santello2019}. A growing number of arguments are accumulating in favor of the theory of continuity and joint coordinated activity of neuron-astrocyte functional networks \citep{Perea2014_1, Kastanenka2019, Kofuji2021}. In the tripartite synapses \citep{Halassa2007,Perea2009} the astrocyte serves as the third part modulating the synaptic transmission. 

A biologically plausible computational model of working memory implemented by a spiking neuron network (SNN) interacting with a network of astrocytes was first proposed in our recent work \citep{Gordleeva2021} and later by \citep{DePitt2021}. The astrocytes operate via calcium transients at a much slower time scale of a few seconds by releasing gliotransmitters that modulate synaptic transmission in neurons and, hence, their firing rate. The working memory is associated with item-specific patterns of astrocyte-induced enhancement of synaptic transmission in neuronal networks. 

Our work \citep{Gordleeva2021}, as the majority of conceptual and mathematical models of neuronal memory, operates with \textit{binary information}. But the real-world data are analogous, not binary. The exploitation of binary information patterns in neural networks was the consequence of ''digitizing'' neuronal signals which are naturally continuous and possess analogous characteristics, which change gradually, e.g. firing rate, timings, phase.  The ''black-and-white'' (BW) paradigm can be easily enhanced to ''colored'' (CL) in artificial digital systems by simple ''spatial'' scaling by increasing the number of bits. The situation is different in the neuronal systems targeting brain-inspired processing when there are no chances for such a scaling. The transition from BW to CL dynamics will require conceptual changes in the models. When recognizing non-binary (grayscale or color) images, stimuli should be converted into signals of spiking neurons. For example, in recent papers \citep{Kulkarni2018,Wozniak2020} sensory neurons are under the influence proportional to the intensity of the corresponding pixels. Other studies have proposed SNNs for grayscale \citep{Lee2018,Yu2021} and color \citep{Cao2014} image recognition. However, such SNNs belong to the class of convolutional networks composed of a hierarchy of stacked convolutional layers. Training of parameters is carried out in order to contrast the boundaries of objects, which are clearly expressed just in binary images. Thus, the circuit processing SNN input signal should contain an algorithm for translating the input image into neural instructions, or the network should have a complex artificial architecture. These factors limit the biological relevance of the models.

Synaptic plasticity represents directed changes of synaptic weights either facilitating or depressing particular connections. In terms of information encoding, such changes are binary, and their main function is the BW representation of the memorized information.  The revealed dependence of the level of calcium elevations generated by astrocytes on the neural activity \citep{Bindocci2017} allows astrocytes to be involved in the regulation of synaptic transmission \citep{Araque2014}. This modulation is gradual and can provide proportional control of the connection efficacy. In other words, analogous information encoding can be possible due to the astrocytes.               

In this paper, we employ our bioinspired model of SNN accompanied by astrocytes \citep{Gordleeva2021} and show how it can reliably store ''colored'' information for several seconds. To the best of the authors’ knowledge, this is the first time that a spiking neuron-astrocyte network has been shown to be able to implement a robust analogous memory, which can be used in brain-inspired artificial intelligence frameworks. For illustration, we take greyscale images as the information patterns and encode them into the level of input currents of the neuronal layer. Due to the interaction with the astrocyte layer, the patterns can be further stored in the network and maintained during the characteristic time interval of the astrocyte activation, e.g. several seconds. During this time the patterns can be retrieved if an appropriate image, e.g. close to the original, comes to the input. After that, the pattern completely disappears and the network becomes ready to store another image.

\section{Colored memory and image recognition in the neuron-astrocyte network model}

\begin{figure}
\centering
\includegraphics[width=1\columnwidth]{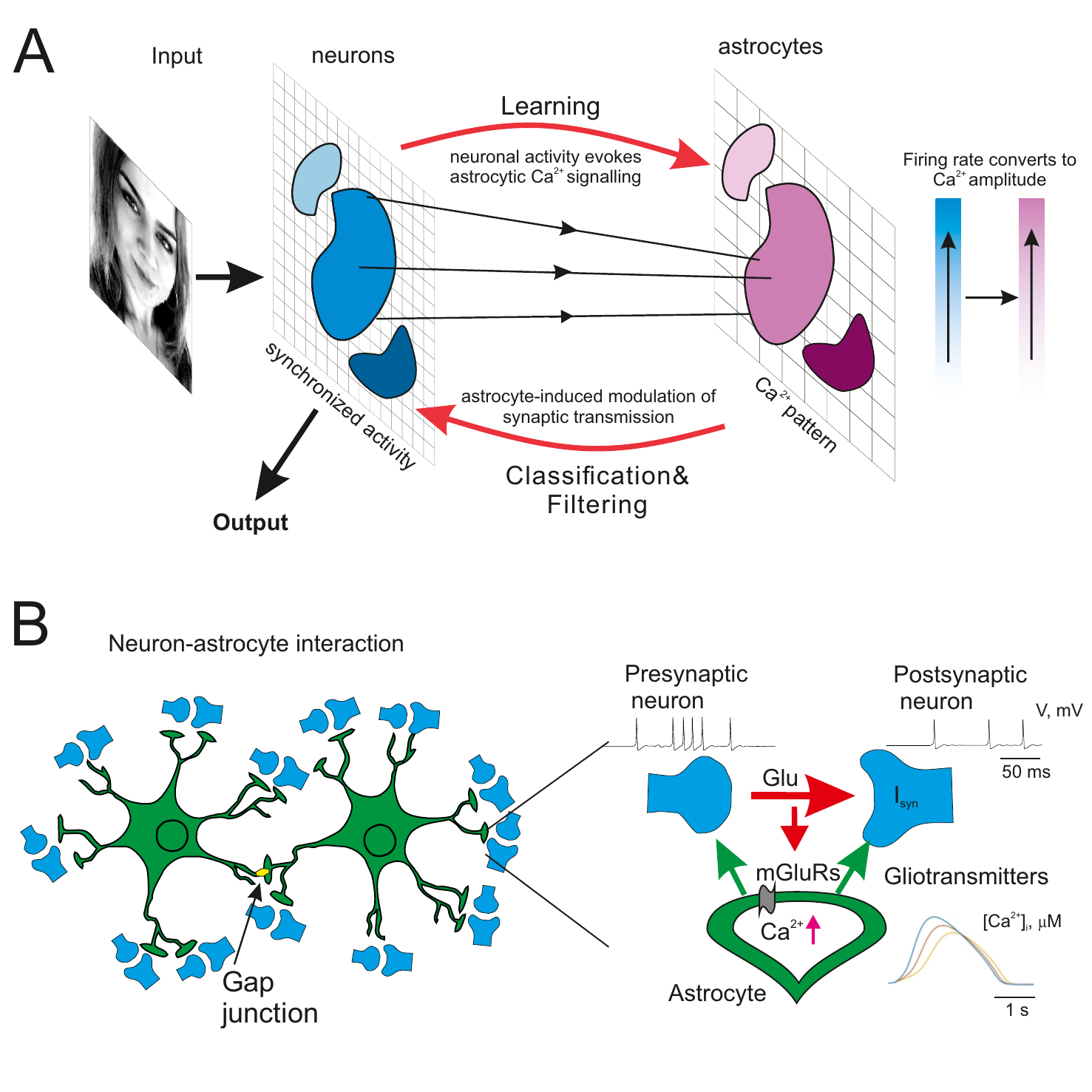}
\caption{(a) System concept and neuron-astrocyte network architecture. The input signal is fed to the layer of neurons. Shades of blue indicate the firing rates of corresponding neurons. Neurons and astrocytes interact bidirectionally: each astrocyte interconnected with a neuronal ensemble of $N_a=16$ neurons with  $4\times4$ dimensions with overlapping in one row. Neuronal spiking rate converts to [Ca$^{2+}$] amplitude, in turn, astrocytes modulate synaptic transmission. The output pattern is decoded as a mean neuronal firing rate. (b) Neuron-astrocyte interaction. The synchronized activity in the neuronal ensemble triggers the elevation of intracellular Ca$^{2+}$ concentration in astrocytes. The global events of Ca$^{2+}$ elevation in astrocytes result in glutamate release, which can modulate the synaptic strength of all synapses corresponding to the morphological territory of a given astrocyte. We consider that the astrocytic glutamate-induced potentiation of the synapse consists in NMDAR-dependent postsynaptic slow inward currents (SICs) generation \citep{Fellin2004,Chen2012} and mGluR-dependent heterosynaptic facilitation of presynaptic glutamate release \citep{Perea2007,Navarrete2008,Navarrete2010}.}\label{fig_topo}
\end{figure}

The neuron-astrocyte network has two interconnected layers: the SNN and the astrocytic network. The SNN composes of randomly sparsely connected excitatory Izhikevich's neurons \citep{Izhikevich2003} with non-plastic synapses arranged in a two-dimensional layer. This layer is interconnected with the astrocytic layer modeled by Ullah's model \citep{ULLAH2006} with local inter-cellular diffusive connections. Each astrocyte bidirectionally communicates with ensembles of $N_a$ neurons. Astrocytes are activated by coordinated activity of the neighboring neurons, e.g. when an input is applied to the neuronal layer. Astrocytic calcium activation induces gliotransmitter release, which modulates the synaptic transmission in neuronal ensemble corresponding to the astrocyte. Such astrocyte-induced synaptic regulation results in the formation of spatially distributed clusters of synchronized neurons. The temporal and amplitude characteristics of astrocytic feedback are determined by its calcium dynamics. This biological relevant mechanism of bidirectional coordination between neuronal and astrocytic activities provides loading, storage, and retrieval of information patterns in the proposed model. The neuron-astrocyte network architecture is schematically illustrated in Fig.~\ref{fig_topo}. Detailed description of the model construction and parameter meaning can be found in our previous paper \citep{Gordleeva2021}. Key mathematical details are summarized in \ref{sec_details}. 

\begin{figure}
	\centering
	\includegraphics[width=1\columnwidth]{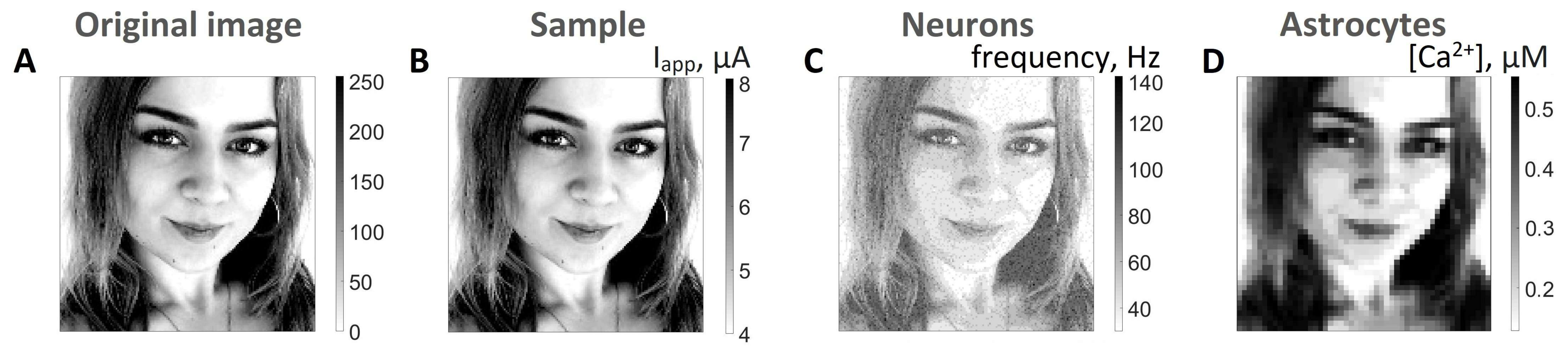}
	\caption{(a) The original image $I$ in 256 shades (8-bit image: values from 0 to 255), (b) the amplitudes of the input currents $I^{(i,j)}_{\text{app}}$, applied to the neuronal layer, (c) the mean neuronal firing rate in the network during the presentation of the sample pattern, (d) intracellular Ca$^{2+}$ concentrations in the astrocytic layer.}\label{fig_Iapp}
\end{figure}

We trained the network to memorize grayscale images. The original 8-bit image (Fig.~\ref{fig_Iapp}a) was converted to the pattern of input current, $I^{(i,j)}_{\text{app}}$, (Fig.~\ref{fig_Iapp}b) and fed to the neuronal layer. Stimulation protocol description can be found in \ref{sec_stimul}. In response to these signals, the neurons fire at different rates depending on the amplitude of the input current (Fig.~\ref{fig_Iapp}c). Differences in the activity of neural ensembles lead to a variety of Ca$^{2+}$ events in astrocytes interacting with it. Fig.~\ref{fig_Iapp}d shows the Ca$^{2+}$ pattern formed in the astrocytic layer. Such sample-specific distribution of Ca$^{2+}$ concentration in the astrocytic layer lasts for several seconds.

\begin{figure}
	\centering
	\includegraphics[width=1\columnwidth]{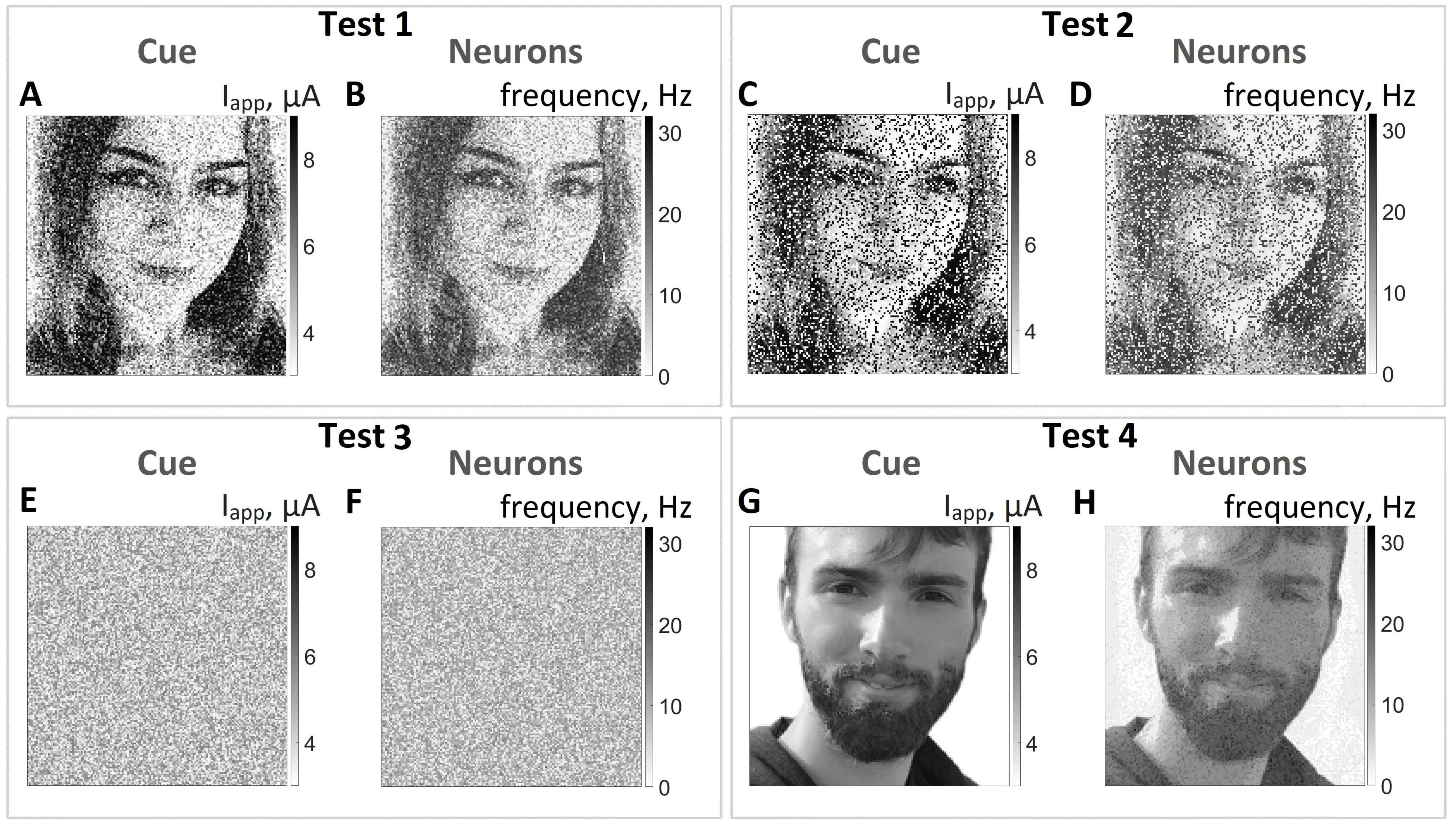}
	\caption{Snapshots of testing the neuron-astrocyte network without modulation of synaptic transmission by astrocytes. (a,c,e,g) are the testing images. (a) is the sample image distorted by 80{\%} Gaussian noise; (c) is the sample image distorted by 40{\%} "salt and pepper" noise; (e) uniform noise; (g) a new image. (b,d,f,h) are the neural network cued recalls. The mean neuronal firing rate in a time window of 500 ms from the beginning of the test image presentation is shown.}\label{fig_tests_no_astro}
\end{figure}

\begin{figure}
	\centering
	\includegraphics[width=1\columnwidth]{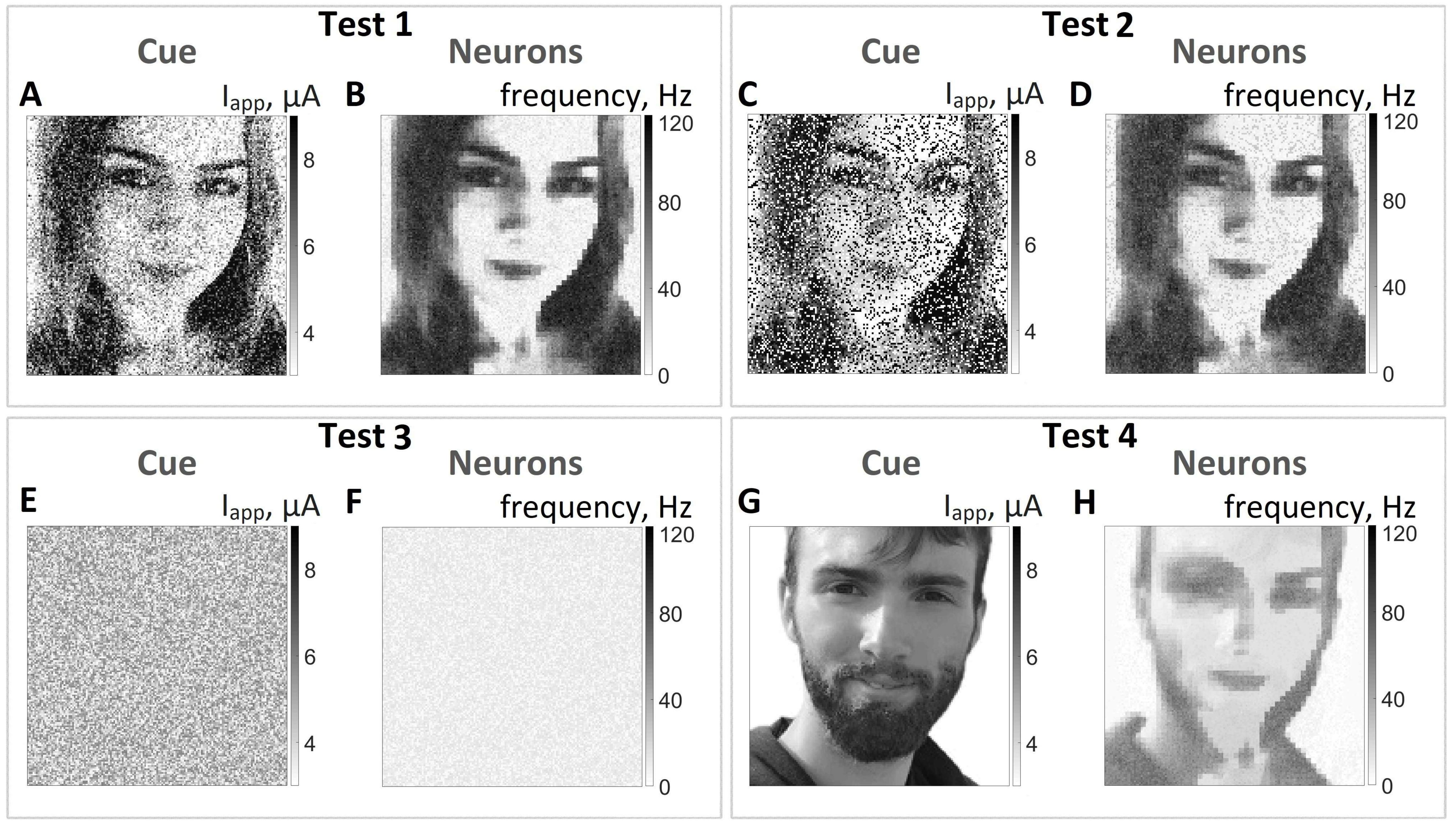}
	\caption{Snapshots of testing neuron-astrocyte network. (a,c,e,g) are  the testing input signals. (a) is the sample image distorted by 80{\%} Gaussian noise; (c) is the sample image distorted by 40{\%} "salt and pepper" noise; (e) uniform noise; (g) a new image.  (b,d,f,h) are the neural network cued recalls. The mean neuronal firing rate in a time window of 500 ms from the beginning of the test image  presentation is shown.}\label{fig_tests}
\end{figure}

We estimate the learning performance of the proposed neuron-astrocyte network model by an image recognition problem. For this purpose, we used four test images: the sample image distorted by 80{\%} Gaussian noise (Fig.~\ref{fig_tests_no_astro}a), by 40{\%} salt{\&}pepper noise (Fig.~\ref{fig_tests_no_astro}c), uniform noise (Fig.~\ref{fig_tests_no_astro}e), and a new image (Fig.~\ref{fig_tests_no_astro}g). To illustrate the impact of astrocytes in the image classification task performed by neuron-astrocyte network, we compared the system recalls with and without astrocytic modulation of synaptic transmission. Fig.~\ref{fig_tests_no_astro} shows that the neuronal layer working on its own can only repeat the input signal without information processing. The results of four tests performed by the full neuron-astrocyte network model are demonstrated in Fig.~\ref{fig_tests}. Fig.~\ref{fig_tests}(a,c,e,g) contain four types of input test images and Fig.~\ref{fig_tests}(b,d,f,h) represent the system recalls shown as the mean neuronal firing rate distributions. The proposed neuron-astrocyte network model can recognize and effectively restore the distorted test image. In the first and second tests, in which the network was fed the noisy matching image, our system significantly reduced additional noise Fig.~\ref{fig_tests}(b,d). Applying noise (Fig.~\ref{fig_tests}e) or nonmatching test image (Fig.~\ref{fig_tests}g) to the neuron-astrocyte network results in a nonspecific (Fig.~\ref{fig_tests}f) or chimera-like (Fig.~\ref{fig_tests}h) output. 

\begin{figure}
	\centering
	\includegraphics[width=1\columnwidth]{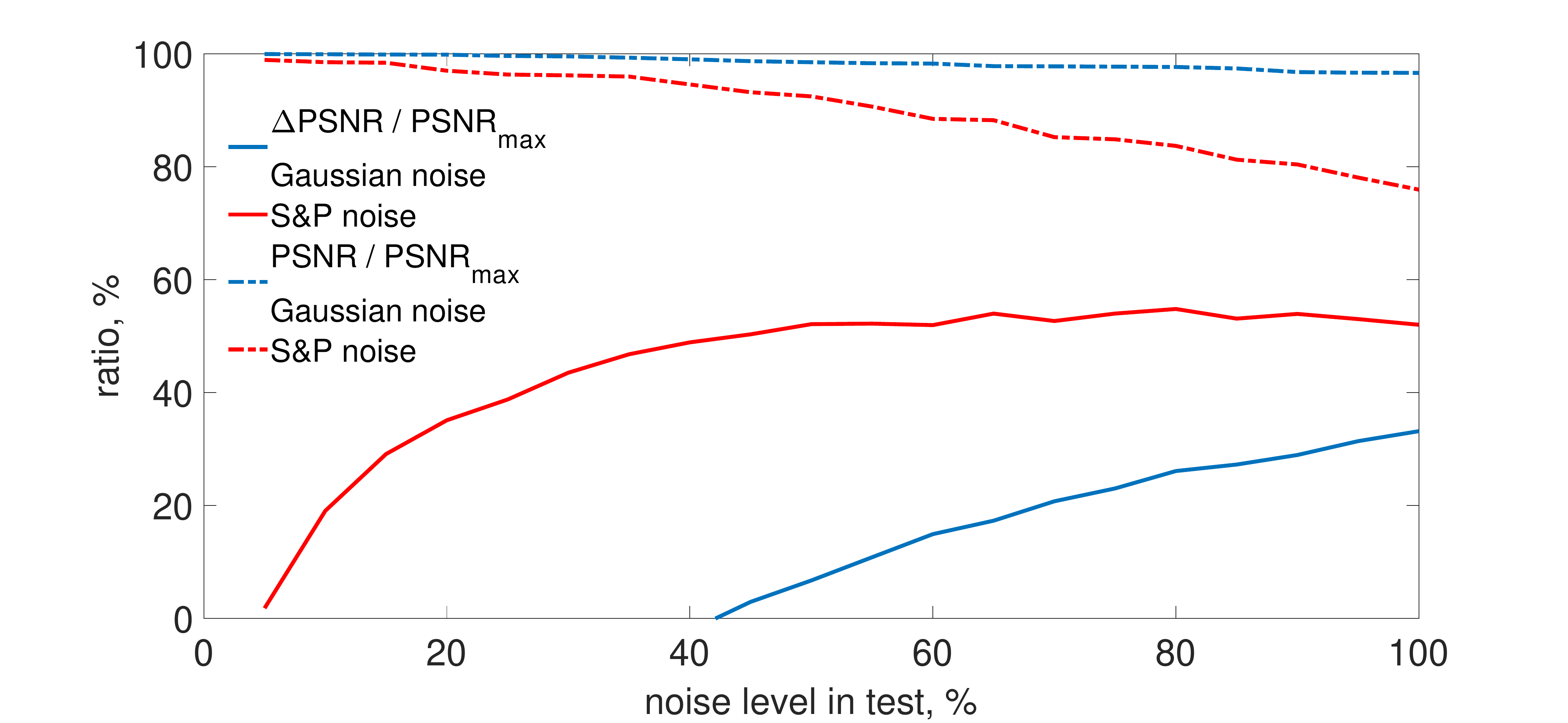}
	\caption{The neuron-astrocyte network model robustness to noise. The dependencies of the PSNR of model recall on the noise level. The dashed lines are the PSNR of model recall related to the maximum PSNR value. The solid lines correspond to the PSNR improvement of the test image in system recall. The blue and red correspond to the Gaussian noise and salt{\&}pepper types of noise, respectively. }\label{fig_noise}
\end{figure}

\begin{figure}
	\centering
	\includegraphics[width=1\columnwidth]{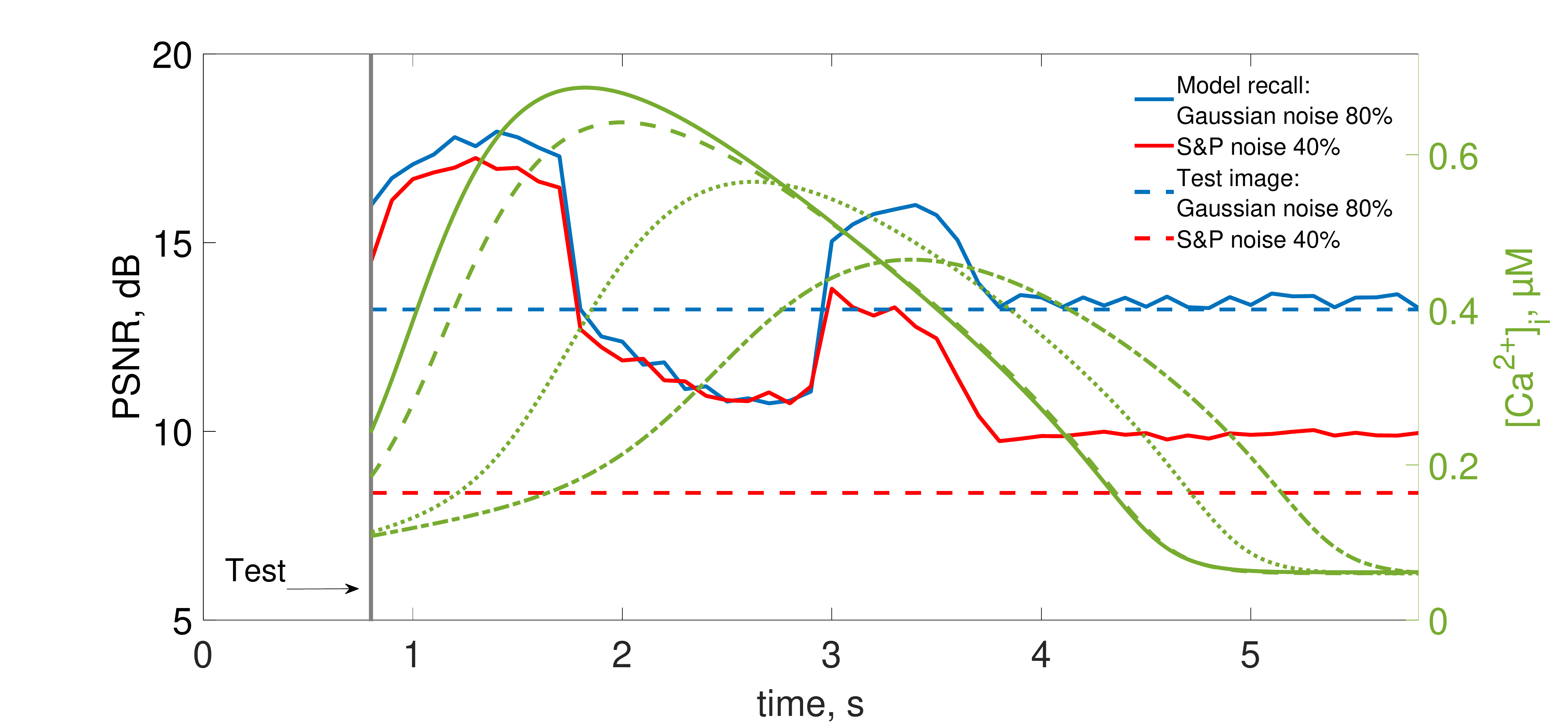}
	\caption{The PSNR of model recall and corresponding astrocytic activity in time. The time corresponds to the moments when the test image was presented. The blue and red curves correspond to the PSNRs of the model recalls in response to test images distorted by 80{\%} Gaussian noise and by 40 {\%} salt{\&}pepper noise, respectively.  Examples of calcium activity in astrocytes are shown in green. The dashed lines are the PSNR of test images.}\label{fig_PSNR}
\end{figure}

To characterize robustness to noise of the proposed neuron-astrocyte network model, we investigated the quality of model retrieval depending on the noise level in the test image. We use two different types of random noise: salt{\&}pepper impulse noise and Gaussian white noise. We examined the ability of our model to remove and reduce noise in an image. In the case of the pulse noise, the noise pixels could be either 1 or 0, which makes them significantly different from image pixels, and the neuron firing rates are significantly different from the neuronal ensemble, respectively. When the noise level is not high, the neuronal correlated activity evokes the astrocyte-mediated feedback which can decrease or increase the firing rates of noise neurons. For Gaussian noise, all pixels of the image change their intensity depending on the noise level. In this case, the astrocyte-induced regulation of synaptic weights restores the general level of activity and the synchronization in the neural ensemble. We measure the PSNR between the recalled pattern (e.g. Fig.~\ref{fig_tests}(b,d)) and the ideal sample image (see section \ref{sec_metric}) as conventional quality metric of image processing systems. Please note that the maximum possible recall PSNR$_{max}$ to the response on the ideal image in the system is 18.295 dB (which is not a very large value) because the resolution of our system has been determined by the radius of the interaction of astrocytes with neurons. The results are well illustrated in Fig.~\ref{fig_noise} and Table~\ref{tab_PSNR}.  The PSNR in {\%} denotes the PSNR of recalls related to the PSNR$_{max}$. We can see that the neuron-astrocyte network can robustly retrieve the memorized image even for a high noise level. The model significantly improved the PSNR for pulse noise for all values within its level and for Gaussian noise for large values of its intensity (Fig.~\ref{fig_noise}). The high level of pulse noise destroys coordinated activity in the neural ensembles which prevents astrocyte-mediated synaptic modulation and, as a result, disturbs the retrieval of formation. Calcium patterns in the astrocytic layer are not frozen and their dynamics is determined by the intracellular biophysical mechanisms. Therefore, the astrocyte-induced feedback and the system recall that depends on it will vary in time. To investigate this, we apply a test image to the system at different time moments corresponding to different distribution schemes of calcium pattern amplitudes in the astrocytic layer. Fig.~\ref{fig_PSNR} shows how the PSNR recall depends on the astrocytic calcium dynamics. A larger difference between the amplitudes of calcium impulses in astrocytes leads to an increase in the difference between the activity levels of neural ensembles and thus to the recall bit depth enhancement and the recall quality improvement.

\begin{table}
      \centering
      \caption{The PSNR recalls (in dB) for different noise levels in test image (mean $\pm$ standard deviation for 10 tests)}\label{tab_PSNR}
      \begin{tabular}{ p{0.25\columnwidth}  p{0.13\columnwidth}  p{0.13\columnwidth} p{0.13\columnwidth} p{0.13\columnwidth} p{0.13\columnwidth}} 
          \\
          noise level: & 20 {\%} & 40 {\%} & 60 {\%} & 80 {\%} & 100{\%}\\ 
          \hline 
         Gaussian noise: \\
         test image & $24.2\pm0.04$ & $18.53\pm0.05$ & $15.34\pm0.03$ & $13.18\pm0.05$ & $11.63\pm0.06$ \\
         model recall & $18.27\pm0.11$ & $18.12\pm0.24$ & $17.98\pm0.17$ & $17.87\pm0.08$ & $17.68\pm0.07$ \\
         
         model recall {\%} & $99.86\pm0.11$ & $99.04\pm0.24$ & $98.28\pm0.17$ &	$97.68\pm0.08$ & $96.64\pm0.07$\\
         \hline
         salt{\&}pepper noise: \\
         test image & $11.38\pm0.11$ & $8.37\pm0.06$ & $6.61\pm0.04$ & $5.36\pm0.02$ & $4.38\pm0.02$ \\
         model recall & $17.75\pm0.21$ & $17.1\pm0.13$ &	$16.26\pm0.04$ & $15.23\pm0.06$ & $13.88\pm0.09$ \\
         model recall {\%} & $97.02\pm0.22$ & $93.47\pm0.14$ & $88.88\pm0.05$ &	$83.25\pm0.07$ & $75.87\pm0.12$\\

\hline           
      \end{tabular}
     \end{table}

\section{Discussion}

We have shown how astrocytes accompanying neuronal synaptic connections can enhance the possibility of the neuronal network to store and retrieve gradual (analogous) information patterns. Greyscale images were used to stimulate our two-layer neuron-astrocyte network. Corresponding synchronous activation of the astrocytic layer allows the system to store images in the form of levels of astrocyte calcium signal during the characteristic duration of calcium transients. Furthermore, different levels of calcium were associated with different strengths of modulation of the synaptic connections in the neuronal layer. Consequently, in the neuronal layer, the images have appeared in the form of activity patterns with different firing rates. During the storage interval, the system maintained the information and could retrieve it if the appropriate clue was shown in the input. We show that the retrieval was quite effective even if a noisy clue pattern was shown.

The role of the astrocytes in brain information processing has been intensively debated in neuroscience in recent years \citep{Kastanenka2019}. By modulating  synaptic transmission, they can be involved in many computational functions of the brain circuits \citep{Santello2019, Kofuji2021}. Today, we have a variety of experimental facts indicating a similar functional role of astrocytes and neurons in perception processes, for example in the processing of visual stimuli. Along with metabolic, homeostatic, and other supporting functions \citep{deHoz2016}, Muller glia cells in the retina provide the delivery of visual information – light, from the anterior surface of the retina to photoreceptors with minimal losses \citep{Franze2007}. Muller cells participate in the structural organization of the retina by creating non-overlapping microdomains that integrate through gap-junctions \citep{Ramirez1996}. This organization allows glial subnets to communicate over long distances \citep{Oberheim2009}. It was shown that astrocytes, like neurons, generate calcium signals in response to visual stimuli, with distinct spatial receptive fields and sharp tuning to a visual stimulus \citep{Schummers2008, Sonoda2018}. Wherein a significant overlap of the receptive fields of astrocytes and nearby neuronal cells was revealed \citep{Schummers2008}. Interestingly, it was found recently that sensory stimulation evokes astrocytic calcium signals with similar temporal dynamics to neurons \citep{Stobart2018}. At variance with neuronal activations, the astrocyte calcium transients are gradual in amplitude \citep{Semyanov2020}. These features indicate that the astrocytes can supply digitized neuronal computations by an analogous component that can significantly increase the computational power of brain circuits. 

The presented result indicates that the spiking neuron-astrocyte network can provide robust analogous information encoding due to the astrocytic modulation of synaptic transmission mechanisms. This is a small but important step in ongoing research on the development of the brain-inspired artificial intelligence. Practically, the performance, for example, in terms, of the accuracy of neuromorphic computing implemented by spiking neuronal networks is still behind modern deep-learning networks in most learning tasks \citep{Shakirov2018}. The main reason for the intensified ongoing research efforts in designing brain-like hardware systems that implement neuronal and synaptic computations through spike-driven communication besides the understanding of brain mechanisms is that it can enable energy-efficient machine intelligence \citep{Roy2019}. It is believed that exploitation of spatio-temporal encoding in SNNs results in a more efficient exchange of information. In this regard, the experimentally and theoretically revealed ability of the astrocytes to evoke the local spatial synchronization in neuronal ensembles due to the activity-dependent short-term synaptic plasticity can become a promising additional feature of training algorithms for SNNs. Another important point that should be stressed is that short-term memory implemented by astrocytes is characterized by one-shot learning and is maintained during the interval of slow astrocytic calcium dynamics. Including the astrocyte-mediated synaptic plasticity in SNN learning algorithms can help to achieve better results than deep learning with the challenge of training on fewer data.

\section{Acknowledgments}

This research was supported by the RFBR projects No. 19-32-60051, 20-32-70081.

\appendix
\section{Model detalization}\label{sec_details}
\subsection{Spiking neuron-astrocyte network model}
 The neuron-astrocyte network consists of two layers:  the spiking neuronal network with dimension $W\times H$ ($151\times 151$) and the astrocytic network. The SNN consists of Izhikevich neurons \citep{Izhikevich2003} connected by random excitatory synaptic connections.  The astrocytic network is $M\times N$ ($50\times 50$) square lattice with only nearest-neighbor connectivity. The dynamics of the intracellular calcium concentration in each astrocyte is described by the Ullah model \citep{ULLAH2006}. A bidirectional neuron-astrocyte interaction was modeled. Each astrocyte interacts with $N_a=16$ neurons located spatially close to it. A graphical representation of the network topology is shown in Fig. ~\ref{fig_topo}. The model was integrated using the $4^{th}$ order Runge-Kutta method with a time step of 0.1 ms. All parameters used in this computational study are given in Table 1 (neuron-astrocyte network parameters) and our previous paper \citep{Gordleeva2021}. The code is available at https://github.com/altergot/neuro-astro-network-grayscale.

\subsection{Neuronal network}
\label{sec_neur_model}
The Izhikevich neuron \citep{Izhikevich2003} was chosen as a model to describe the dynamics of each neuron in our network due to its biological relevance and computational efficiency. This model is described by the following differential equations \citep{Izhikevich2003}:

\begin{equation}
\label{eq:Izh_main}
\begin{aligned}
&\frac{dV^{(i,j)}}{dt} = 0.04V^{(i,j)^{(2)}} + 5V^{(i,j)} - U^{(i,j)} + 140 +I^{(i,j)}_{\text{app}} + I^{(i,j)}_\text{syn}; \\
&\frac{dU^{(i,j)}}{dt} = a ( b V^{(i,j)} - U^{(i,j)});\\
\end{aligned}
\end{equation}
with the auxiliary after-spike resetting:
\begin{equation}
    \label{eq:Izh_cond}
    \text{if } V^{(i,j)} \ge \text{30 mV, then}
        \begin{cases}
            V^{(i,j)}\gets c \\U^{(i,j)} \gets U^{(i,j)} + d,\\
        \end{cases}
\end{equation}
where $i,j$ $(i = \overline{1,W}, j = \overline{1,H})$ are the neural indices, $V$ is the transmembrane potential, $t$ is the time in ms. $I^{(i,j)}_{\text{app}}$ is the input signal. $I^{(i,j)}_{\text{syn}}$ is total synaptic current from all presynaptic neurons $N^{(i,j)}_{\text{in}}$, which is calculated as follows \citep{Kazantsev2011,Esir2018}:

\begin{equation}
    \label{eq:I_syn}
    \begin{aligned}
    &I^{(i,j)}_{\text{syn}} =\sum_{k=1}^{N^{(i,j)}_{in}} \frac{g^{(i,j)}_{\text{syn}}(E_{\text{syn}}-V^{(i,j)})}{1+\exp(\frac{-V^{k}_{\text{pre}}}{k_{\text{syn}}})};\\
    \end{aligned}
\end{equation}
where the parameter $g^{(i,j)}_{\text{syn}}$ is the synaptic weight: $g^{(i,j)}_{\text{syn}}=\eta+\nu^{(m,n)}_{Ca}$, $\eta$ is the weight of the synaptic connection, $\nu^{(m,n)}_{Ca}$ is the astrocyte-induced modulation of the synaptic weight (see section: "Bidirectional neuron-astrocyte interaction"). $E_{\text{syn}}=0$ is the synaptic reversal potential for excitatory synapses. $V_{\text{pre}}$ is the membrane potential of the presynaptic neuron, $k_{\text{syn}}$ is the slope of the synaptic activation function. In this model, we do not take into account synaptic and axonal delays.

The architecture of synaptic connections between neurons is random: for each neuron, the number of output connections is fixed and equal to $N_{\text{out}}$. Thus, the probabilities of the formation of a local and remote synaptic connection are the same.

First, we tested the functioning of our model with the same weights of synaptic connections between all neurons in the neuron-astrocytic network. Differences in the total synaptic input current resulted in some noise in the firing rate response when the original training image was fed. To reduce this effect, at the beginning of the session, we pre-train synaptic connections depending on the shades of the training image $I$:

\begin{equation}
\label{eq:eta}
\begin{aligned}
\eta = 0.9 ^ {|I^{(i,j)}-I^{(i^{*},j^{*})}|} * (\eta_{max}-\eta_{min})+\eta_{min}, 
\end{aligned}
\end{equation}
where $I^{(i,j)}$ is the pixel shade value of the training source image $I$ from the interval [0; 255] corresponding to the presynaptic neuron ($i$,$j$), $I^{(i^{*},j^{*})}$ is the pixel shade value of the training source image $I$ corresponding to the postsynaptic neuron $(i^{*},j^{*})$. Thus, a small difference in the shades of the pixels of the original training image corresponding to the presynaptic and postsynaptic neurons corresponds to a strong synaptic connection between this pair of neurons. The greater the difference in pixel shades, the weaker the synaptic connection between the corresponding neurons.

\subsection{Astrocytic network}
\label{sec_astro_model}
Astrocytic dynamics is determined by changes in the concentration of two main substances: inositol 1,4,5-triphosphate (IP$_3$) and intracellular calcium (Ca$^{2+}$). The main astrocytic intracellular calcium store is the endoplasmic reticulum (ER). Ca$^{2+}$ can be released from the ER through the membrane channels into the cytoplasm, which corresponds to an increase in intracellular calcium concentration. Ca$^{2+}$ flux from the ER to the cytoplasm, $J_{\text{$ER$}}$, is a non-linear function of calcium concentration [Ca$^{2+}$] and is controlled by the IP$_3$ concentration. The rate of this flow is determined by the fraction of channels on the ER membrane that are in the open (non-inactivated) state $h$. The reverse flow of calcium $J_{\text{$pump$}}$ from the cytoplasm to the ER is an active transport that pumps calcium back into the ER and is directed conversely to the concentration gradient.

To describe the dynamics of the intracellular [Ca$^{2+}$] in each astrocyte ($m$, $n$) of our network, we used the Ullah model \citep{ULLAH2006}, which qualitatively reflects the main features of the calcium dynamics of astrocyte (for more details about this model and the biophysical meaning of all flows and parameters, see \citep{ULLAH2006}). This model consists of the following differential equations:

\begin{equation}
\label{eq:Ullah}
\begin{aligned}
&\frac{d[Ca^{2+}]^{(m,n)}}{dt}  = J_{\text{ER}}^{(m,n)}-J_{\text{pump}}^{(m,n)}+J_{\text{leak}}^{(m,n)}+J_{\text{in}}^{(m,n)}-J_{\text{out}}^{(m,n)}+\text{diff}^{(m,n)}_{Ca};\\
&\frac{dh^{(m,n)}}{dt}  = a_2 \left(d_2\frac{IP_3^{(m,n)}+d_1}{IP_3^{(m,n)}+d_3}(1-h^{m,n})-[Ca^{2+}]^{(m,n)}h^{(m,n)} \right);\\
&\frac{d[IP_3^{(m,n)}]}{dt}  = \frac{IP_3^*-IP_3^{(m,n)}}{\tau_{IP3}}+J_{\text{PLC$\delta$}}^{(m,n)}+J_{\text{glu}}^{(m,n)}+\text{diff}^{(m,n)}_{IP3}\\
\end{aligned}
\end{equation}
where $J_{\text{$leak$}}$ is the leakage flux from the ER to the cytosol. The fluxes $J_{\text{$in$}}$ and $J_{\text{$out$}}$ describe the exchange of calcium with the extracellular space, $m$,$n$ ($m$ = 1,\ldots, $M$, $n$ = 1,\ldots,$N$) are the astrocyte indices. The parameter [IP$_3^*$] denotes the steady-state concentration of IP$_3$, $J_{\text{PLC$\delta$}}$ describes the production of IP$_3$ by phospholipase C$\delta$ (PLC$\delta$), $J_{\text{$glu$}}$ describes the glutamate-induced IP$_3$ production in response to neural activity. The fluxes are expressed as follows:

\begin{equation}
\label{eq:flux}
 \begin{aligned}
&J_{\text{ER}} = c_1v_1[Ca^{2+}]^3h^3IP_3^3\frac{(c_0/c_1-(1+1/c_1)[Ca^{2+}])}{((IP_3+d_1)([Ca^{2+}]+d_5))^3};\\
&J_{\text{pump}} = \frac{v_3[Ca^{2+}]^2}{k_3^2+[Ca^{2+}]^2};\\
&J_{\text{leak}} = c_1v_2(c_0/c_1-(1+1/c_1)[Ca^{2+}]);\\
&J_{\text{in}} = \frac{v_6IP_3^2}{k_2^2+IP_3^2};\\
&J_{\text{out}} = k_1[Ca^{2+}];\\
&J_{\text{PLC}\delta} = \frac{v_4([Ca^{2+}]+(1-\alpha)k_4)}{[Ca^{2+}]+k_4}
\end{aligned} 
\end{equation}

Astrocytes formed networks by connecting through gap-junctions Cx43 \citep{Yamamoto1990,Nagy2000,Nimmerjahn2004}. Diffusion currents of IP$_3$ molecules and Ca$^{2+}$ ions, $\text{diff}_{Ca}$ and $\text{diff}_{IP_3}$, can be expressed as follows:

\begin{equation}
\label{eq:astro_dif}
\begin{aligned}
&\text{diff}^{(m,n)}_{Ca} = d_{Ca}(\Delta [Ca^{2+}])^{(m,n)};\\
&\text{diff}^{(m,n)}_{IP3} = d_{IP3}(\Delta IP_3)^{(m,n)};\\
\end{aligned}
\end{equation}

where $d_{Ca}$ and $d_{IP_3}$ describe the Ca$^{2+}$ and IP$_3$ diffusion rates, respectively. In our model each astrocyte is coupled with only four nearest neighbors. $(\Delta [Ca^{2+}])^{(m,n)}$ and $(\Delta IP_3)^{(m,n)}$ are the discrete Laplace operators:  

\begin{equation}
\label{eq:astro_Laplace}
\begin{aligned}
&(\Delta [Ca^{2+}])^{(m,n)}&=([Ca^{2+}]^{(m+1,n)}+[Ca^{2+}]^{(m-1,n)}+[Ca^{2+}]^{(m,n+1)}+[Ca^{2+}]^{(m,n-1)}-4[Ca^{2+}]^{(m,n)});\\
&(\Delta IP_3)^{(m,n)}&=((\Delta IP_3)^{(m+1,n)} + (\Delta IP_3)^{(m-1,n)}+(\Delta IP_3)^{(m,n+1)}+(\Delta IP_3)^{(m,n-1)}-4(\Delta IP_3)^{(m,n)}).\\
\end{aligned}
\end{equation}

\subsection{Bidirectional neuron-astrocyte interaction}
\label{sec_neuro_astro_interaction}
Each astrocyte in the spiking neuron-astrocyte network interacts with a 4 by 4 ensemble of $N_a$ neurons with overlapping in one row. The spiking activity of neurons leads to the release of the neurotransmitter glutamate $G$ from the presynaptic terminal into the synaptic gap. The amount of $G$ that reached the astrocyte is described by the following equation: \citep{Gordleeva2012,Pankratova2019}:
\begin{equation}
\label{eq:Glu}
\begin{aligned}
\frac{dG^{(i,j)}}{dt} & = -\alpha_\text{glu} G^{(i,j)}+k_\text{glu}\Theta(V^{(i,j)}-30 mV),\\
\end{aligned}
\end{equation}

where $\alpha_\text{glu}$ is the glutamate clearance constant, $k\text{glu}$ is the release efficiency, $\Theta$ is the Heaviside step function, and $V^{(i,j)}$ is the membrane potential of a neuron ($i$,$j$).
Glutamate contacts metabotropic glutamate receptors (mGluR) on the astrocyte membrane and initiates the production of IP$_3$. The $J_{\text{$glu$}}$ variable in the equation describes glutamate-induced IP$_3$ production and is modeled as:
\begin{equation}
    \label{eq:Jglu}
    \begin{aligned}
    J_{\text{glu}} = \begin{cases}
                      G_\text{sum}^{(m,n)}, \qquad &\text{if} \quad G_\text{sum}^{(m,n)} > G_{thr1},  \\
                        0, \qquad &\text{otherwise};\\
                     \end{cases}
    \end{aligned}
\end{equation}
where $G_{\text{$thr1$}}$ is the threshold for the total amount of glutamate $G$ released by all neurons associated with the astrocyte ($m$,$n$). $G_\text{sum}^{(m,n)}$ is total glutamate $G$ that reached an astrocyte ($m$,$n$):

\begin{equation}
\label{eq:sum_glu1}
\begin{aligned}
G_\text{sum}^{(m,n)} = \left(\sum_{(i,j)\in{N_a}}G^{(i,j)}\right), 
\end{aligned}
\end{equation}

Higher neuronal activity causes more glutamate to be released. This, in turn, leads to a longer duration and greater amplitude of the $J_{\text{$glu$}}$ elevation. Differences in the $J_{\text{$glu$}}$ elevations initiated by the activity of neural ensembles lead to differences in Ca$^{2+}$ dynamics of astrocytes corresponding to these neurons through IP$_3$ production. Thus, the larger the amplitude and duration of the $J_{\text{$glu$}}$ elevation, the longer and higher-amplitude calcium event it will cause.

The proposed model of spiking neuron-astrocyte network takes into account the following mechanisms of astrocytic  enhancement of excitatory synaptic transmission due to the gliotransmitter action. Astrocytic glutamate-induced (i) potentiation of the synapse through the generation of the slow inward currents (SICs) in the postsynapse \citep{Chen2012, Fellin2004}; and (ii) mGluR-dependent heterosynaptic facilitation of presynaptic glutamate release \citep{Perea2007,Navarrete2008, Navarrete2010}. The revealed dependence of the level of calcium elevations generated by astrocytes on neural activity allows astrocytes to gradually regulate synaptic transmission \citep{Araque2014}. For simplicity, the relationship between the astrocyte Ca$^{2+}$ concentration and synaptic weight of the affected synapses $g_{syn}$, is described as follows:
\begin{equation}
\label{eq:nu_Ca}
\begin{aligned}
\nu^{(m,n)}_{Ca} = 
\nu_{Ca}^*\frac{[Ca^{2+}]^{(m,n)}-[Ca^{2+}]_{thr}}{[Ca^{2+}]_{max}}\Theta([Ca^{2+}]^{(m,n)}-[Ca^{2+}]_{thr});\\
\end{aligned}
\end{equation}
where $\nu_{Ca}^*$ is the strength of the astrocyte-induced modulation of the synaptic weight, $\Theta$ is the Heaviside step function, $[Ca^{2+}]_{max}$ is the maximum Ca$^{2+}$ concentration in astrocytic layer at the specific moment. Feedback from astrocytes to neurons is activated when $[Ca^{2+}]$ is greater than $[Ca^{2+}]_{thr}$, and the total amount of glutamate released by the neurons corresponding to the astrocyte is greater than  the threshold: $G_\text{sum}^{(m,n)}>G_{thr2}$. The duration of synaptic transmission by astrocytes is fixed and equal to $\tau_\text{astro}$ according to the experimental data of astrocyte-induced SICs dynamics \citep{Fellin2004}.

\subsection{Stimulation protocol}
\label{sec_stimul}
The size of each visual stimulus is equal to the neural network size: $W$ × $H$. The original image $I$ was quantized in 256 shades (8-bit image: values from 0 to 255) (Fig.~\ref{fig_Iapp}a). Then, to train the network, for each of the 256 shades, a value was assigned from a range of linearly spaced values from 4 to 8. (Fig. ~\ref{fig_Iapp}b). Each pixel value was used as the amplitude of the input signal $I^{(i,j)}_{\text{app}}$ from equation~(\ref{eq:Izh_main}) for the corresponding neuron ($i$,$j$). Thus, the input signal $I^{(i,j)}_{\text{app}}$ for a neuron ($i$,$j$) was a rectangular pulse with an amplitude $A_{\text{stim}}$ equal to the pixel ($i$,$j$) value and duration $t_{\text{stim}}$. 

To illustrate how the network can store and retrieve greyscale patterns, we used four images: the same photo with pixel intensities normalized to the range [4; 9] and with an additional 80 {\%} Gaussian noise (Fig.~\ref{fig_tests_no_astro}a), the same photo with pixel intensities normalized to the range [4; 9] and with an additional 40 {\%} "salt and pepper" noise (Fig.~\ref{fig_tests_no_astro}c), uniform noise with values from the range [4; 9] (Fig.~\ref{fig_tests_no_astro}e), another photo with pixel intensities normalized to the range [4; 9] (Fig.~\ref{fig_tests_no_astro}g). Test images were also presented as an input signal to neurons with the duration $t_{\text{test}}$.

The salt{\&}pepper noise level (in $\%$) is the fraction of noisy pixels.
The Gaussian noise level (in $\%$) represents the ratio of standard deviations of the white Gaussian noise and of the whole normalized image.

\begin{table}
      \centering
      \caption{Neuron-astrocyte network parameters}\label{tab_neuro1}
      \begin{tabular}{ p{0.15\columnwidth}  p{0.55\columnwidth}  p{0.15\columnwidth} } 
          \\
          Parameter & Parameter description & Value  \\ 
          \hline 
         $W\times H$ & neural network grid size & $151\times 151$ \\
            $\eta_{max}$ & maximum pre-trained weight of synaptic connection without astrocytic influence  & 0.025  \\
            $\eta_{min}$ & minimum pre-trained weight of synaptic connection without astrocytic influence  & 0.001  \\
            $E_{\text{syn}}$ & synaptic reversal potential for excitatory synapses & 0 mV  \\
            $k_{\text{syn}}$ & slope of synaptic activation function & 0.2 mV  \\
            $N_{\text{out}}$ & number of output connections per each neuron & 100  \\
            $M\times N$ & astrocytic network grid size & $50\times 50$  \\
            $d_{Ca}$ & Ca$^{2+}$ diffusion rate & 0.05 s$^{-1}$ \\
            $d_{IP_3}$ & IP$_3$ diffusion rate & 0.05 s$^{-1}$ \\
            $N_a$ & number of neurons interacting with one astrocyte & 16, $4\times 4$  \\ 
            $\alpha_\text{glu}$  & glutamate clearance constant & 10 s$^{-1}$  \\ 
            $k_\text{glu}$ & efficacy of the glutamate release & 600 $\mu$M s$^{-1}$ \\
            $G_\text{thr1}$ & threshold concentration of glutamate for IP$_3$ production & 2  \\
            $G_\text{thr2}$ & threshold of total glutamate required for the occurrence of astrocytic modulation of synaptic transmission & 3  \\
            $\nu_{Ca}^*$ & strength of astrocyte-induced modulation of synaptic weight & 0.1  \\
            $[Ca^{2+}]_\text{thr}$ & threshold concentration of Ca$^{2+}$ for astrocytic modulation of synapse & 0.2 $\mu$M \\
            $\tau_\text{astro}$ & duration of astrocyte-induced modulation of synapse & 300 ms  \\
            $t_\text{stim}$ & stimulation duration & 100 ms \\
            $t_\text{test}$ & cue stimulation length & 100 ms \\

\hline           
      \end{tabular}
     \end{table}

\subsection{Metrics for evaluating retrieval quality} 
\label{sec_metric}
To assess the retrieval quality of the developed neuron-astrocyte network, we used the PSNR method:
\begin{equation}
\label{eq:PSNR}
\begin{aligned}
&PSNR = 10\log_{10}\frac{MAX_{I}^{2}}{MSE},\\
&MSE = \frac{1}{WH}\sum_{i=1}^{W}\sum_{j=1}^{H}{[I(i,j)-K(i,j)]}^{2},\\
\end{aligned}
\end{equation}
where $MAX_{I}$ = 255 is the maximum possible pixel value. To use this method, we converted all the results obtained (mean neuronal firing rate during testing) into 8-bit greyscale images $K$ and compared them with the original image $I$. We calculated the mean firing rate of each neuron during testing as the mean number of spikes in a time window of 500 ms from the beginning of the test image presentation.

\bibliography{refs}





\end{document}